\documentclass[journal=jpcbfk,manuscript=article]{achemso}
\setkeys{acs}{articletitle = true}
\usepackage[version=3]{mhchem} 
\usepackage[T1]{fontenc}       

\author{Ivan Brihuega}
\author{Felix Yndurain}
\email{felix.yndurain@uam.es.}
\affiliation{Departamento de F\'{i}sica de la Materia Condensada
and Condensed Matter Physics Center (IFIMAC).
Universidad Aut\'{o}noma de Madrid. Cantoblanco. 28049 Madrid.
Spain.}

\date{\today}

\title{Selective Hydrogen Adsoprtion in Graphene Rotated Bilayers}

\begin{document}

\begin{abstract}
The absorption energy of atomic hydrogen at rotated graphene bilayers is studied using ab initio methods based on the density functional theory including van der Waals interactions. We find that, due to the surface corrugation induced by the underneath rotated layer and the perturbation of the electronic density of states near the Fermi energy, the atoms with an almost AA stacking are the preferential ones for hydrogen chemisorption.  The adsorption energy difference between different atoms can be as large as 80 meV. In addition, we find that, due to the logarithmic van Hove singularities in the electronic density of states at energies close to the Dirac point,  the adsorption energy of either electron or hole doped samples is substantially increased. We also find that the adsorption energy increases with the decrease of the rotated angle between the layers. Finally, the large zero point energy of the C-H bond (${\sim 0.3 eV}$) suggests adsorption and desorption of atomic hydrogen and deuterium should behave differently.
\end{abstract}

\maketitle

\section{I. Introduction}
Different point defects like vacancies and impurities in graphite and graphene have been largely considered in the past to induce new behaviors like semiconducting \cite{Semiconducting}, stiffness \cite{Gomez-Navarro}, magnetism \cite{Pereira,LopezSancho,Yazyev, Yazyev-Review, Ugeda, Yo-solo, Yo-Science, Tucek} and many other properties \cite{RevModPhys, PointDefects, Pykal}. Among the different point defects, hydrogen adsorption on graphitic surfaces has been extensively studied in different areas such as interstellar chemistry, hydrogen storage and plasma/fusion physics. Since the apparition of graphene the field has gained new interest because, as recently proven, the adsorption of a single hydrogen atom leads to the emergence of a graphene magnetic moment of 1 Bohr magneton  \cite{Yo-Science}. Single atomic hydrogen has been studied both theoretically and experimentally.  It is characterized as forming a well defined covalent bond with one carbon atom which is distorted from the planar graphene structure forming a puckered structure close to the  ideal $sp^{3}$ one. The atomic hydrogen magnetic moment is transferred to the carbon atoms near the C-H bond. There is a general agreement on this picture (see Gonz\'{a}lez Herrero et al.\cite{Yo-Science} and references therein). The adsorption energy, i.e. the energy difference between hydrogen being far away from graphene and being chemisorbed, is of the order of 1 eV. There are however some open questions, in particular the tendency of H atoms to form non-magnetic clusters in the surface of graphite and graphene, which hinders the possibility to induce a large magnetic moment in graphene by the simple adsorption of a large number of H atoms. It is therefore crucial to find strategies to modify the energetic landscape seen by H atoms on graphene surfaces, with the specific goal of increasing the adsorption energies. Our work shows that graphene rotated bilayers are an ideal playground for such a goal. The combination of atomic corrugation and the existence of van Hove singularities provides a large adsorption energy difference between the different C atoms of the graphene layer. Moreover, the adsorption energy increases with the decrease of the rotated angle between the layers. Our calculations also proof that electron or hole doping is an extremely effective way to increase the adsorption energies and points to Deuterium as a clever choice to increase further more the stability.
\begin{figure}[ht]
\includegraphics[width=100mm]
{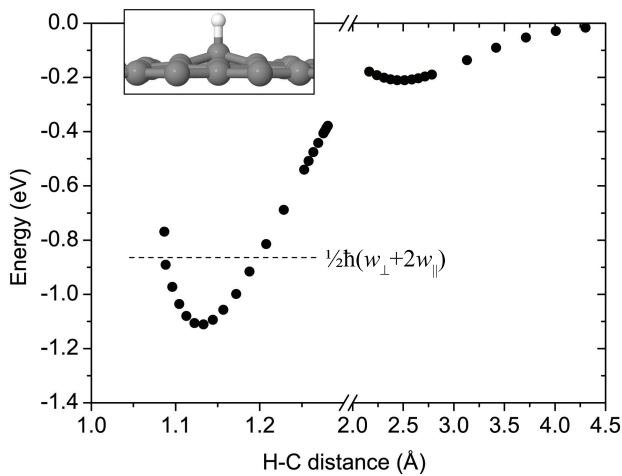}
\caption{Total energy (in eV) of  a hydrogen atom adsorbed at a graphene monolayer versus the 
hydrogen-carbon bond length in {\AA}. The energy is referred to the case of hydrogen and graphene being far apart.  The horizontal line indicates approximately the calculated zero point energy of the C-H vibration modes (energy with respect to the well minimum).}
\label{H-Adsorption}
\end{figure}

\section{II. Computational Details}
In order to study the geometrical and electronic structure of hydrogen adsorption in rotated graphene bilayers we perform first principles density functional \cite {DFT1, DFT2} calculations using the SIESTA code \cite {Siesta1, Siesta2} which uses localized orbitals as basis functions \cite{Orbitals}. We use a double $\zeta$ basis set (in some instances the results are checked increasing the basis with polarized orbitals), non-local norm conserving pseudopotentials and for the exchange correlation functional we use the the generalized gradient approximation (GGA)\cite{GGA} including van der Waals interaction in the way implemented by Roman-Perez and Soler \cite{Guillermo-Roman} with the functional developed by Berland and Hyldgaard \cite{vdW-BH}. In some cases the results are compared with those obtained with the functional originally developed by Dion et al. \cite{vdW-DRSLL}. The results depend quantitatively on the van der Waals implementation but qualitative are very similar (see bellow). The calculations are performed with stringent criteria in the electronic structure convergence (down to $10^{-5}$ in the density matrix), 2D Brillouin zone  sampling (up to 600 $k$-points), real space grid (energy cut-off of 400 Ryd) and equilibrium geometry (residual forces lower than $2\times10^{-2}$ eV/\AA). Due to the rapid variation of the density of states at the Fermi level, we used a polynomial smearing method \cite{smearing}.  
To study defects we use the super cell approximations in the way that we end up with an interaction between defects in the repeated unit cell.  The effect of this interaction has been checked in some cases increasing the size of the unit cell. To define and characterize the unit cells of the rotated bilayers we follow the notation of 
Trambly de Laissardi\`{e}re et al. \cite{Moire}; if we call $\vec{a_{1}}$ and  $\vec{a_{2}}$ the primitive lattice vectors, the new ones for the (n$\_$m) unit cell are: $\vec{t}=n \vec{a_{1}}+m\vec{a_{2}}$ and $\vec{t'}=-m \vec{a_{1}}+(n+m)\vec{a_{2}}$. In this way the new unit cell is rotated an angle $\theta$  with respect to the conventional primitive cell. This rotation angle is given by
$cos\theta =\frac{n^{2}+4nm+m^{2}}{2(n^{2}+nm+m^{2})}$,
and in a rotated (n$\_$m) unit cell the number of atoms is $2(n^2+m^2+nm)$ per layer.

In graphene, it is generally accepted that point defects like adsorbed atoms or single vacancies unbalance the  number of atoms in both sublattices giving rise therefore to a localized magnetic moment. This localized moment is associated to a new state not present in perfect graphene and localized at the Dirac point energy. The character of this state is peculiar  \cite{Besenbacher-1} and has been studied and characterized extensively in model hamiltonians  \cite{Satpathy, Pereira, RevModPhys}. It decays like $1/r$ as it has been proved experimentally in graphite \cite{Ugeda}.  Beyond model hamiltonians many first principles calculations been performed and there is a large list of published papers. To mention some of them: hydrogen and vacancies have been studied by Yazyev \cite{Yazyev-Review, Yazyev} , fluorine has been studied for instance by Sofo et al.\cite{Balseiro} and Tucek et al.\cite{Tucek} and different $sp^{3}$ defects by Santos et al. \cite{Dani}. 

Before studying hydrogen chemisorption on rotated bilayers, we have looked at the monolayer case. Hydrogen atoms are known to chemisorb on top of carbon atoms \cite{Balseiro, Jeloaica}. In Figure \ref{H-Adsorption} we show a typical curve of the adsorption of hydrogen on a graphene monolayer. The two physisorption and chemisorption regimes are patent in the figure. Similar results have been already reported without van der Waals interaction \cite{Jeloaica, Besenbacher-1, Besenbacher-2, Besenbacher-3} and 
including it \cite{Mohammed}. We have calculated the C-H stretching and bending phonon frequencies such that 
$\hbar\omega _{\perp }=0.317 eV$ 
and
$\hbar\omega _{\parallel }=0.147 eV$
such that the zero point energy is 
$\frac{1}{2}\hbar(\omega _{\perp }+2\omega _{\parallel })=0.305 eV$
in agreement with previous calculations \cite{Herrero} and experiments \cite{CH-vibration}. We can anticipate that similar results are obtained in the bilayer case. It is important to realize that this value is significant as compared with the chemisorption potential well. The expected value for deuterium (0.216 eV) suggests hydrogen and deuterium would show different behavior.  

\section {III. Results}
\subsection{Clean $3\_4$ bilayer}
Most of the calculations in this work have been performed in a 3$\_$4 bilayer (148 carbon atoms altogether) using
the functional of Berland and Hyldgaard \cite{vdW-BH}. 
\begin{figure}[ht]
\includegraphics[width=100mm]
{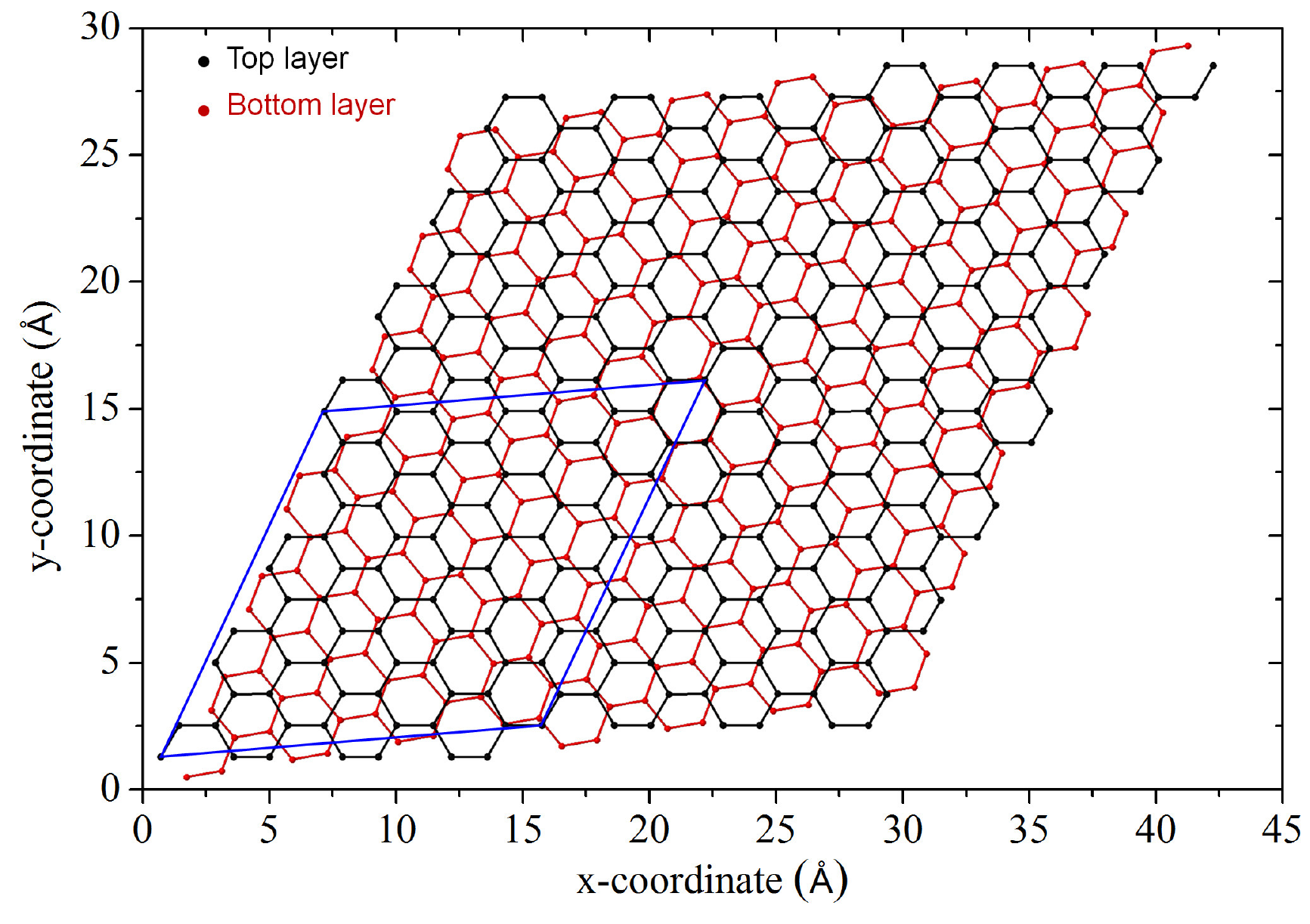}
\caption{(Color online) Atomic structure of a clean 3$\_$4 bilayer (see the main text for notation). The shown unit cell is formed by four primitive cells. Black and red colors distinguish between the two 9.43 degrees rotated layers. The almost AA packing in the center is patent. The primitive unit cell is indicated.}
\label{Moire-3_4-xyz}
\end{figure}
\begin{figure}[ht]
\includegraphics[width=100mm]
{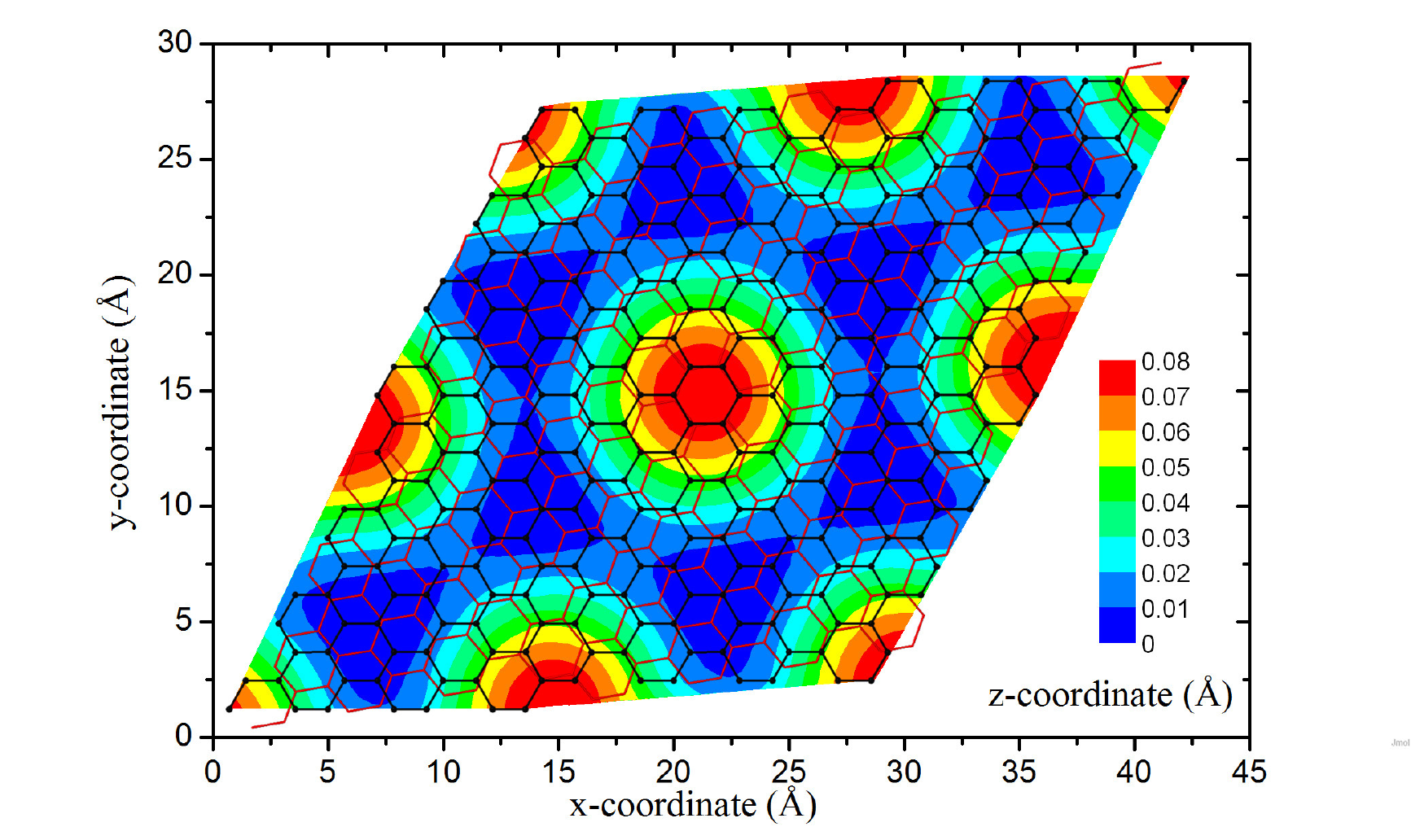}
\caption{(Color online)  Corrugation (in {\AA}) of the upper graphene layer in a fully relaxed 3$\_$4 bilayer. As in Figure \ref{Moire-3_4-xyz} the shown unit cell is formed by four primitive cells. The top/bottom graphene layer is shown in black/red. }
\label{Moire-3_4-Clean}
\end{figure}
We consider the lower layer flat and fixed with the nominal graphene structure
with a minimized nearest neighbor distance of  1.43 \AA, the upper layer is relaxed up to residual forces
of  $2\times10^{-2}$ eV/\AA. After the relaxation, the average separation between the layers is 3.25 \AA. In te super-cell calculation, we consider, in the direction perpendicular to the layers, a vacuum gap of 20 \AA . The atomic configuration of a 2x2 unit cell including four primitive cells
is shown in Figure \ref{Moire-3_4-xyz}. The almost ideal AA packing in the center of the figure is clear.
At this point it is important to study how the top layer is perturbed by the presence of the underneath rotated layer. In Figure \ref{Moire-3_4-Clean} we show the perpendicular coordinate map of the relaxed top layer.
As expected, we first observe how the atoms near the AA configuration are displaced with respect to the plane of the underneath flat layer. For more details see reference \cite{Yo-van-Hove}.

\begin{figure}[ht]
\includegraphics[width=100mm]
{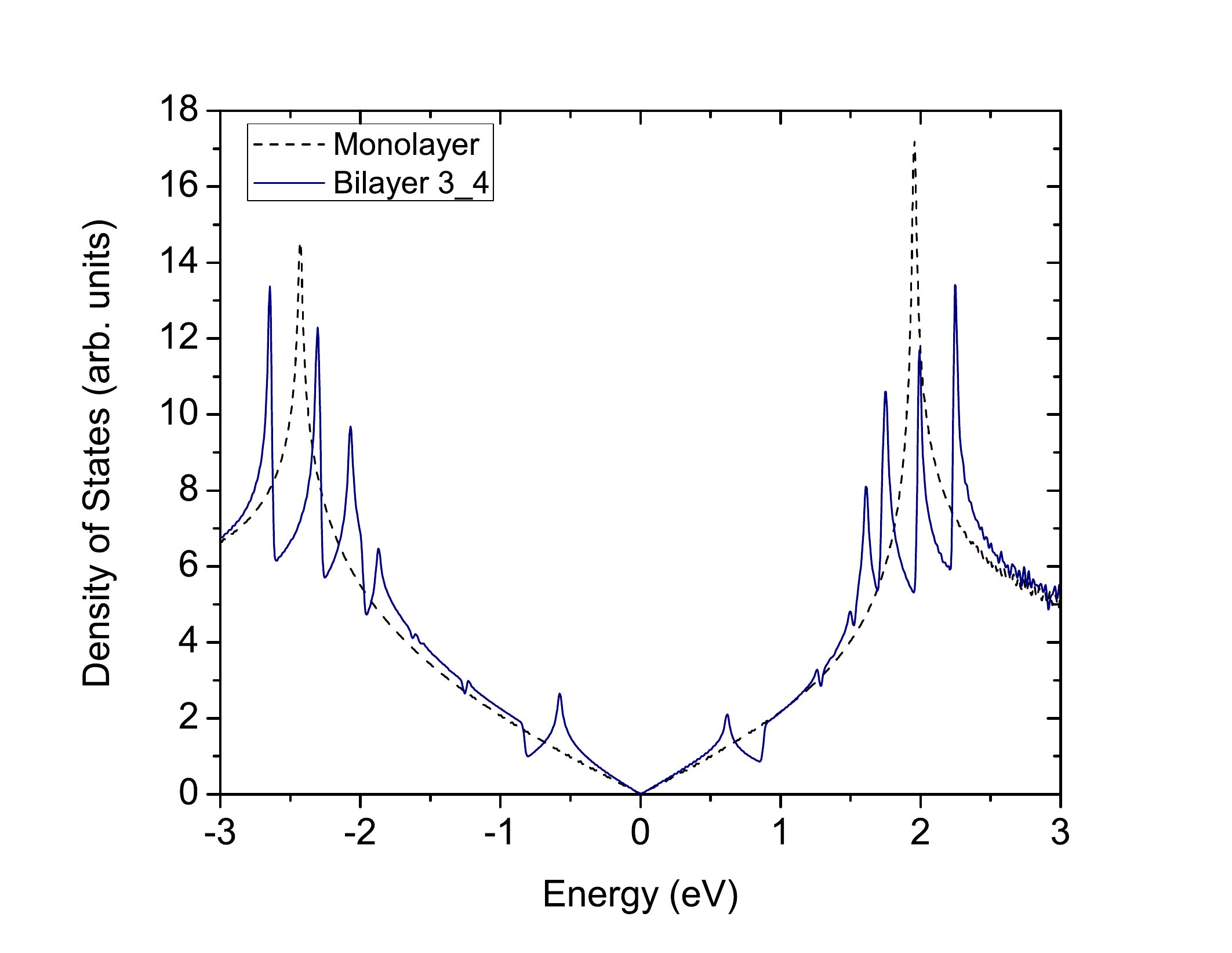}
\caption{(Color online) Density of states (normalized to states per carbon atom) of a graphene monolayer (broken line)
and a 3$\_$4 bilayer (continuous line).}
\label{Moire-3_4.DOS}
\end{figure}

We expect this deviation of the graphene planar structure and the interaction with the underneath layer should perturb the electronic structure. In Figure \ref{Moire-3_4.DOS} we show the total density of states of the relaxed 
bilayer and, for comparison, the density of states of a graphene monolayer. For a better comparison both curves have been normalized to states per carbon atom. We clearly observe several changes on the density due to presence of the bilayer: first, the appearance of the logarithmic singularities peak close to the Dirac point both below and above it. These have been discussed and characterize already in the literature (see reference \cite{Yo-van-Hove} and reference therein). It is relevant at this point to recall that these peaks are closer to the Dirac point the smaller the rotated angle between the layers. Second, there is an important perturbation of the density of states at around 2 eV both above and bellow the Dirac point. In addition, there are new sharp peaks almost evenly separated induced by the new periodicity. This separation depends on the interaction between the layers and can be relevant 
in doped graphene bilayer and in graphene monolayers on top of metals with a Moir\'{e} structure. Third, the density of states close to Dirac point increases with respect to that of graphene due to the presence of the second layer. This can have relevance on the hydrogen adsorption on both clean and doped samples (see below). In addition (not shown) the local density of states depends on the atom we are looking at; for instance the logarithmic peaks are more intense close to the AA region than in the rest of the atoms.    

\subsection{Hydrogen adsorption}
\begin{figure}[ht]
\includegraphics[width=120mm]
{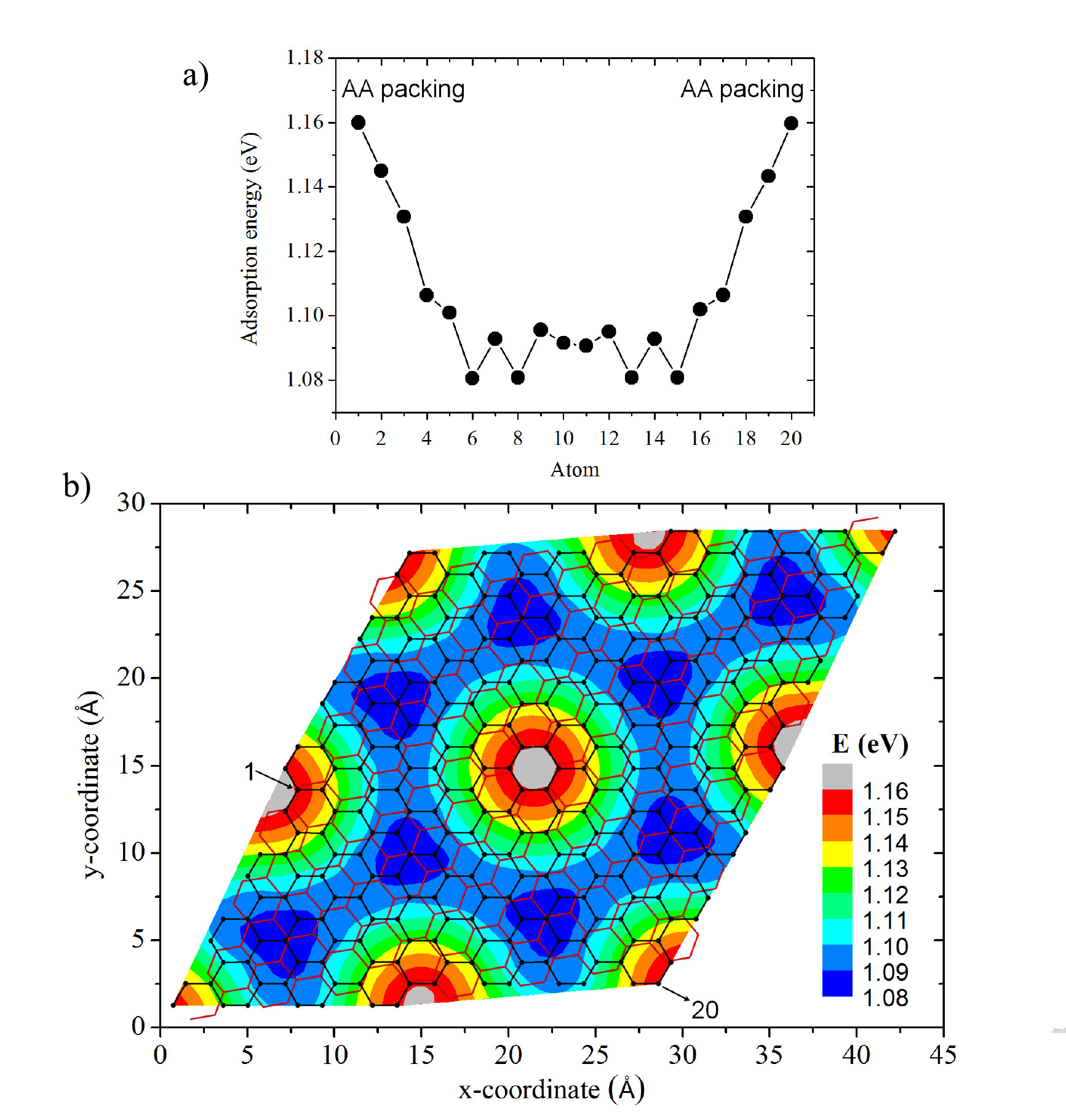}
\caption{Adsorption energy (using the functional of Berland and Hyldgaard \cite{vdW-BH}) at various atoms in the top layer of the $3\_4$ bilayer. In (a) the atoms correspond to a zig-zag chain between the two equivalent atoms 1 and 20, marked by arrows in panel (b), in two quasi AA stacking zones. In (b) the complete adsorption energy map is shown. The top/bottom graphene layer is shown in black/red.}
\label{Ads-different-atoms}
\end{figure}

There have been many calculations of hydrogen adsorbed on graphene (see refs. \cite{Jeloaica, Yazyev, Yo-solo,Yo-Science, Balseiro, Verges, Pantelides, JCP, Buckling} etc.) The calculation of the hydrogen adsorption has indeed to be spin dependent. As discussed previously  \cite{Yo-solo,Yo-Science}, in a non spin-resolved calculation of the electronic structure of a single hydrogen atom there is a very sharp peak in the density of states at the Dirac point close to the Fermi level. The peak is half filled except for a very small (of the order of 0.01 electrons) charge transfer between hydrogen and graphene. Clearly, if the equal spin population constrain is released, a close to 1 $\mu_{B}$ magnetic moment associated to the peak takes place. 
The absorption  may depend on the carbon atom to which hydrogen is bonded to. There are two reasons to expect
the carbon atoms close to the AA packing would be the preferential ones for the hydrogen atom to bind; first, these atoms are pull above the planar configuration being closer to the $sp^{3}$ configuration characteristic of the hydrogen chemisorbed and, second, as indicated above, the electronic density of states close to the Fermi energy is larger on those atoms than in the rest of them. The absorption energy at different atoms in the unit cell is plotted in Figure \ref{Ads-different-atoms} . We immediately observe that adsorption energy is much larger in atoms close to the AA stacking with the underneath layer. The adsorption energy difference between different atoms can be as large as 80 meV. This result clearly indicates that the atoms near a AA stacking with the underneath layer are the preferential ones for hydrogen adsorption. In actual experiments, hydrogen adsorption is commonly performed at RT. Thus, the expected ratio between hydrogen atoms in AA and AB sites ($N_{AB}/N_{AA}$) can be obtained using Boltzmann statistics as $N_{AB}/N_{AA} = exp(E_{(AA-AB)}/kBT)$, where $E_{(AA-AB)}$ is the energy difference between AA and AB adsorption sites. For dilute hydrogen concentrations  $E_{(AA-AB)}\sim 80 meV$, giving $N_{AB}/N_{AA} \sim 22$ at RT. It should be indicated that in the case of the physisorption well, we do not find significant differences between the carbon atoms hydrogen is bonded to. Actually, this well is independent of the xy coordinate \cite{Mohammed}.  

\begin{figure}[ht]
\includegraphics[width=100mm]
{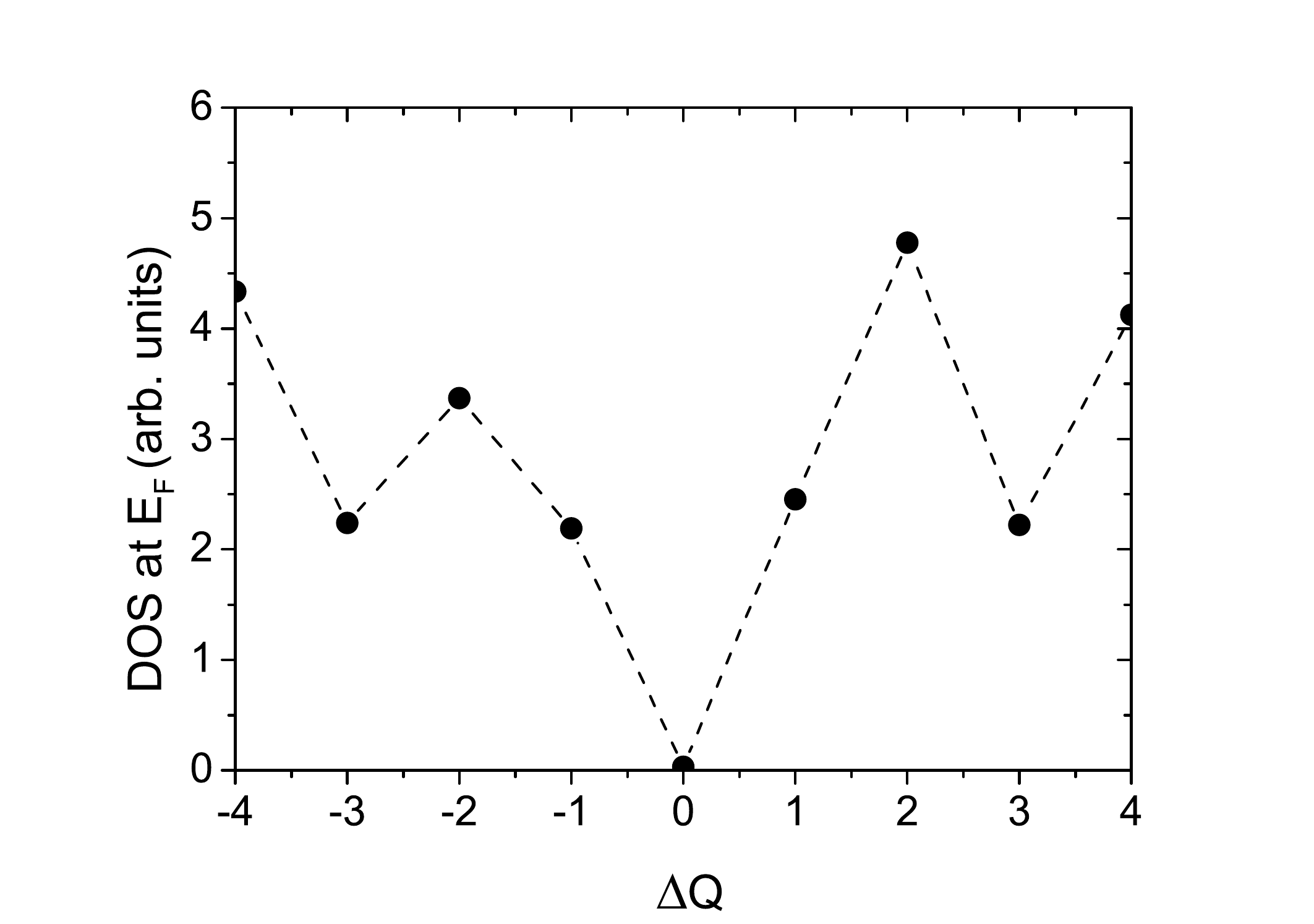}
\caption{(Color online) Variation of the density of states at the Fermi level with doping in the clean fully relaxed $3\_4$ graphene bilayer. The charge $\Delta Q$ refers to an extra or defect of charge to the 592 electron of the undoped case.}
\label{DOS-at-EF-BH}
\end{figure}
\begin{figure}[ht]
\includegraphics[width=100mm]
{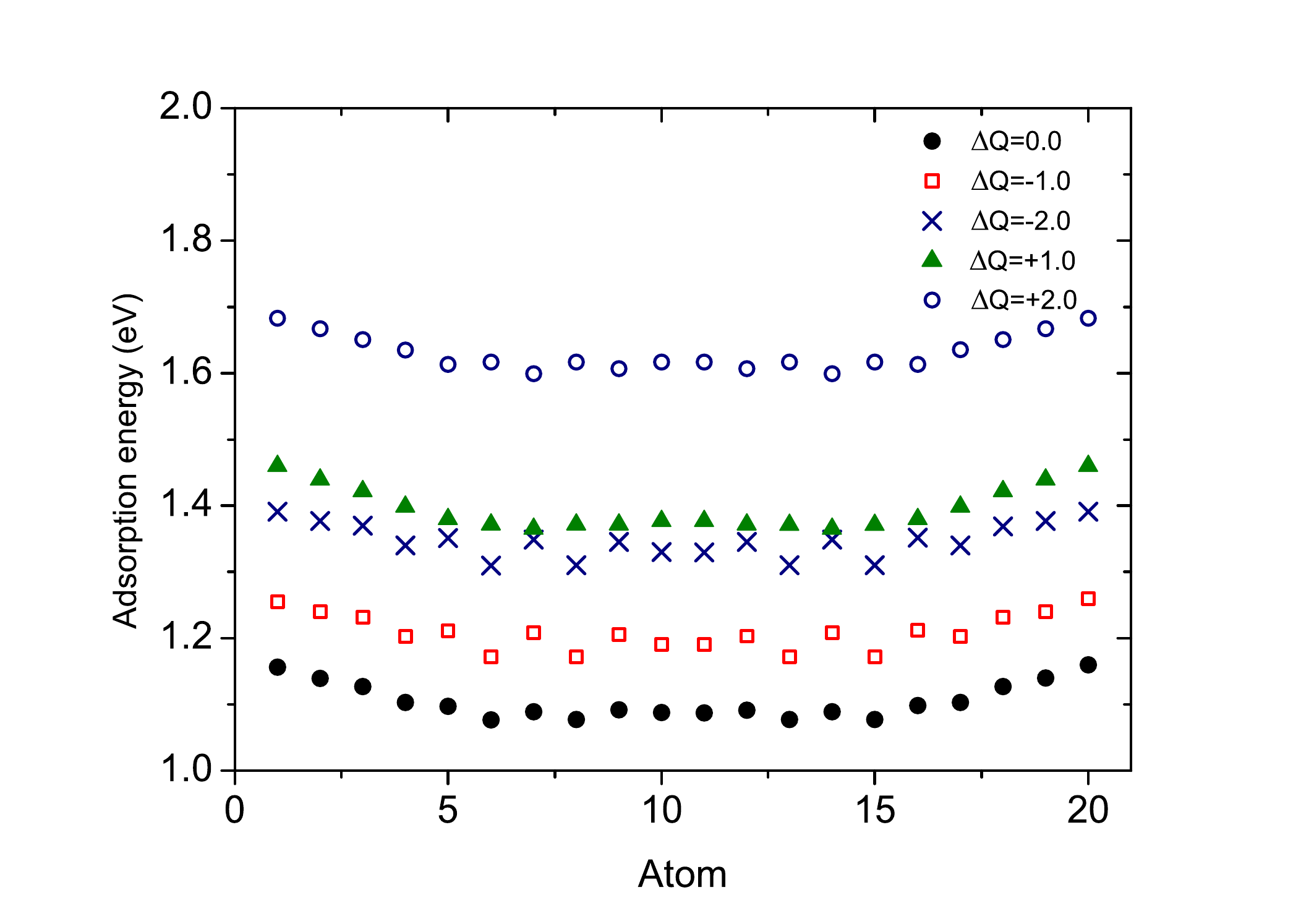}
\caption{(Color online) Adsorption energy at various atoms for different electron ($\Delta Q<0$) and hole ($\Delta Q>0$) doping in the $3\_4$ bilayer. The atoms correspond to a zig-zag chain between two equivalent atoms in two quasi AA stacking zones separated by a primitive lattice vector (see broken line in Figure \ref{Ads-different-atoms}(b))}
\label{Adsorption-vs-doping}
\end{figure}

At this point it is interesting to study the effect of doping on the chemisorption energy \cite{Chinos-doping}. By adding electrons or holes the Fermi energy would move from the Dirac points where the density of states is zero, to energies where the density of states is larger and, eventually, at the van Hove logarithmic singularities, where the density of states is much larger. In  Figure \ref{DOS-at-EF-BH} we show the calculated density of states at the Fermi as a function of the extra charge. The two maxima at around $\sim \pm$ 2 $\Delta Q$ correspond to the Fermi level at the van Hove singularities in the density of states at $\sim \pm$ 0.6 eV (see Figure \ref{Moire-3_4.DOS}). We then expect an increase of the hydrogen adsorption energy. Upon doping, hydrogen, in addition to lose the magnetic moment would find more graphene electronic states to make a stronger bond. Results of the calculations are shown in Figure \ref{Adsorption-vs-doping}. We first observe that the preferential adsorption sites are always at the AA packing irrespective of the doping and, in addition, we observe an important enhancement of the chemisorption energy as we increase either the electron or hole doping.   
Finally we have studied how the chemisorption energy depends on the rotated angle between the layers.
It is known that the smaller the angle the closer are the van Hove singularities to the Dirac point and therefore larger the density of states at the Fermi energy and larger the adsorption energy. We have calculated these energies for
hydrogen chemisorption at preferential carbon atoms near the AA stacking for different stacking angles. Results of the calculations are in Table \ref{Table}. As expected, the adsorption energy increases with the decrease of the angle between the layers. It would be interesting to reduce even more the angle between the layers but, unfortunately, this would increase the size of the unit cell making the calculation unfeasible. 
\begin{table}
  \centering
    \begin{tabular}{| c c c c c c  |}
    \hline
  & Monolayer & 2$\_$3 & 3$\_$4 & 4$\_$5 & 5$\_$6\\ 
\hline
 Angle (degrees) & - & 13.17 & 9.43 & 7.34 & 6.01\\  
 Adsorption energy (eV) & 1.086 & 1.127 & 1.156 & 1.164 & 1.172\\  
       \hline
    \end{tabular}
  \caption{Adsorption energy of atomic hydrogen on top of a monolayer of graphene and at various
  graphene rotated bilayers (see the main text) .  In these cases hydrogen is chemisorbed at the carbon atom with larger chemisorption energy.}
 \label{Table}
\end{table}

\begin{figure}[ht]
\includegraphics[width=100mm]
{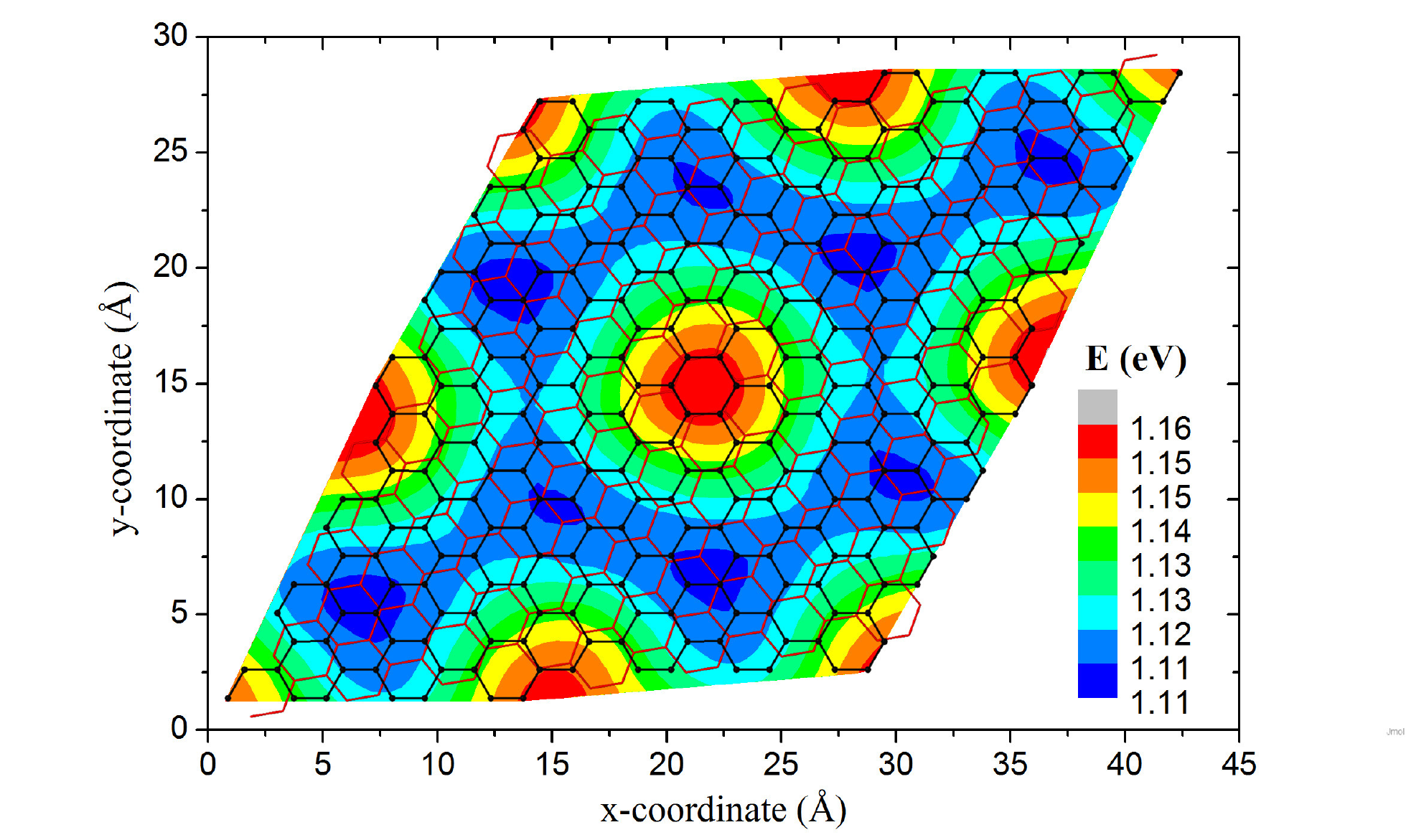}
\caption{(Color online) Adsorption energy of atomic hydrogen chemisorbed on a $3\_4$  graphene bilayer using the functional of Dion et al. \cite{vdW-DRSLL}. As in Figures \ref{Moire-3_4-xyz} and  \ref{Moire-3_4-Clean} the shown unit cell is formed by four primitive cells. The top/bottom graphene layer is shown in black/red.}
\label{Moire-3_4-H-DRSLL}
\end{figure}

We have repeated the calculation of the hydrogen adsorption map using the original Dion et al. \cite{vdW-DRSLL}
van der Waals functional. Results of the calculation are shown in Figure \ref{Moire-3_4-H-DRSLL}. In this case the differences
in adsorption energy at various carbon atoms are not as pronounced as in the case of the  Berland and Hyldgaard \cite{vdW-BH} functional, but, again, the atoms around the AA region are the preferred ones for hydrogen chemisorption.

\subsection{IV. Conclusion}
In summary, we have studied whether or not there are preferential chemisorption sites in rotated graphene bilayers.
From our calculations we can conclude the following: 

\textit{i)} In rotated graphene bilayers the sites close to the AA configuration are the preferential ones for atomic hydrogen chemisorption.

\textit{ii)} The hydrogen physisorption well is independent of the adsorption site.
 
\textit{iii)} Although the preferential carbon sites for hydrogen chemisorption are independent of the sample doping the absolute chemisorption binding energy increases with the doping, either electron or hole.

\textit{iv)} The chemisorption energy increases with the decrease of the angle between the rotated layers.  
 
\textit{v)} Considering the calculated values of the energies of chemisorption and zero point energy of the C-H bond, this has to be considered when analyzing adsorption, desorption and migration of atomic hydrogen in graphene systems. Probably hydrogen and deuterium landscapes are substantially different. Nuclear quantum effects may play an important role in hydrogen adorption \cite{Pykal} and shoul be further investigated.
 
\textit{vi)} It would be interesting to extend this study to other impurities like fluorine and small molecules to elucidate the preferential site adsorption in rotated graphene bilayers. Preliminary calculations with fluorine indicate very similar behavior.


\subsection{Acknowledgements} We would like to thank J. M. G\'{o}mez-Rodr\'{i}guez,  J.J. Palacios and J. M. Soler for  illuminating discussions. I. B. acknowledges MINECO (grants MAT2016-80907-P and PCIN-2015-030), Fundaci\'{o}n Ram\'{o}n Areces and the European Union FLAG-ERA program for financial support. F. Y. acknowledges the  Spanish Ministry of Science and Innovation for financial support through grant FIS2015-64886-C5-5-P.

\providecommand{\latin}[1]{#1}
\makeatletter
\providecommand{\doi}
  {\begingroup\let\do\@makeother\dospecials
  \catcode`\{=1 \catcode`\}=2\doi@aux}
\providecommand{\doi@aux}[1]{\endgroup\texttt{#1}}
\makeatother
\providecommand*\mcitethebibliography{\thebibliography}
\csname @ifundefined\endcsname{endmcitethebibliography}
  {\let\endmcitethebibliography\endthebibliography}{}


\end{document}